\documentclass[prd,twocolumn,showkeys,superscriptaddress]{revtex4}
\usepackage[latin1]{inputenc} 
\usepackage{fontenc}      
\usepackage{graphicx,epsfig,units,color}
\usepackage{amsmath,amssymb}

\newcommand{\bean}{\begin{eqnarray}}
\newcommand{\eean}{\end{eqnarray}}
\newcommand{\be}{\begin{equation}}
\newcommand{\ee}{\end{equation}}

\begin{document}

\title{CMB anisotropies seen by an off-center observer in a spherically symmetric inhomogeneous universe}
\date{\today}

\author{H\aa{}vard Alnes}
\email{alnes@fys.uio.no}
\affiliation{Department of Physics, University of Oslo, PO Box 1048
  Blindern, 0316 Oslo, Norway} 

\author{Morad Amarzguioui}
\email{morad@astro.uio.no}
\affiliation{Institute of Theoretical Astrophysics, University of
  Oslo, PO Box 1029 Blindern, 0315 Oslo, Norway}

\begin{abstract}
The current authors have previously shown that inhomogeneous, but
spherically symmetric  universe models containing only matter can yield a very good fit to the SNIa data and the position of the first CMB peak. In this work we examine how far away from the center of inhomogeneity the observer can be located in these models and still fit the data well. Furthermore, we investigate whether such an off-center location can explain the observed alignment of the lowest multipoles of the CMB map. We find that the observer has to be located within a radius of $\sim\unit[15]{Mpc}$ from the center for the induced dipole to be less than that observed by the COBE satellite. But for such small displacements from the center, the induced quadru- and octopoles turn out to be insufficiently large to explain the alignment.
\end{abstract}
\keywords{inhomogeneous universe models -- dark energy -- observational
  constraints}

\maketitle

\section{Introduction}

In a recent work \cite{alnes:2005}, we studied spherically symmetric inhomogeneous universe models -- the so-called Lema\^itre-Tolman-Bondi (LTB) models. We found that for a certain class of inhomogeneities, such models could easily explain various cosmological observations without introducing dark energy, most notably the luminosity distance-redshift relation of type IA supernovae and the position of the first peak in the CMB spectrum.  The inhomogeneities required are of the form of a spherically symmetric underdense bubble in an otherwise flat and homogeneous Einstein-de Sitter universe, with the observer located at the center of the bubble.


Unless the observer is positioned exactly at the center of the bubble,
the distribution of matter, as seen by the observer, will be anisotropic. This  will affect the observed microwave background and constrain the possible location of the observer, since the CMB dipole must be in agreement  with observations \cite{Bennett:1996ce}. Note that in a homogeneous universe model, this dipole is attributed to the peculiar velocity of the observer. However, as discussed in \cite{Humphreys:1996fd}, in an LTB model there will be an additional contribution to the dipole from the anisotropy of space-time. Thus, the dipole seen by an off-center observer will be due to a combination of kinematic effects and the off-center location.

The anisotropy will also induce higher multipoles in the CMB spectrum.  Moffat~\cite{Moffat:2005jc} proposes this mechanism as a possible explanation for the observed alignment of the CMB quadru- and octopole \cite{deOliveira-Costa:2003pu,Copi:2003kt,Schwarz:2004gk,Ralston:2003pf,Land:2005ad}, since the direction from the observer towards the center of the bubble singles out a ``special'' axis.


In this work we will investigate these induced anisotropies in the CMB to establish how far from the center the observer can be located, and whether they can offer an explanation to the alignment of the lowest multipoles. We find that the observer has to be located within a sphere extending approximately $\unit[15]{Mpc}$ about the origin, in order for the induced dipole to remain within the observed range. However, within this small volume the induced quadru- and octopole turn out to be insufficiently large to explain the alignment.

The paper is organized as follows: In Sect.~\ref{sec:geodesic}, we
deduce the differential equations governing the path and redshift of
photons. We then proceed to work out expressions for the induced
temperature distribution and corresponding CMB multipoles in
Sect.~\ref{sec:temp}. Next, in Sect.~\ref{sec:numeric}, we solve these
equations numerically to find the multipoles as a function of the
position of the observer. This allows us to find LTB models which
agree with both the observed dipole and observations of SNIa and the
position of the first CMB peak, as described in \cite{alnes:2005}. We
present two such models in this section. Finally, in
Sect.~\ref{sec:discussion} we discuss and summarize our work.

\section{The geodesic equation in the LTB space-time}
\label{sec:geodesic}

The spherically symmetric LTB metric is given by \cite{Lemaitre:1933,Tolman:1934,Bondi:1947}
\be
\label{eq:metric}
ds^2 = -dt^2 + \frac{[R'(r,t)]^2}{1+\beta(r)}dr^2+R^2(r,t)d\Omega^2\,, 
\ee
where $R(r,t)$ is a position-dependent scale factor, and $\beta(r)$ is
related to the curvature. 

One of the great advantages of working in this space-time is that the Einstein equations can be solved exactly in the matter-dominated scenario. The function $R(r,t)$ can then be written in terms of a conformal time $\eta$, defined by $\beta^{1/2}dt = Rd\eta$, as
\bean
R &=& \frac{\alpha}{2\beta}(\cosh \eta - 1)\nonumber\\
  &&+R_0\left[\cosh \eta +
  \sqrt{\frac{\alpha+\beta R_0}{\beta R_0}}\sinh \eta\right]\,, \label{eq:defR}\\ 
\sqrt{\beta}t &=& \frac{\alpha}{2\beta}(\sinh \eta-
\eta)\nonumber\\&&
+R_0\left[\sinh \eta + \sqrt{\frac{\alpha+\beta R_0}{\beta 
R_0}}\left(\cosh \eta   -1\right)\right]\,, \label{eq:deft} 
\eean
where $\alpha(r) \geq 0 $ is an arbitrary function, and we have
assumed $\beta(r) > 0$. Furthermore, we have defined $R_0 \equiv R(r,0)$,
with the initial time $t=0$ defined as the time of last scattering.

Photons follow trajectories determined by the geodesic equation,
\be
\frac{d^2 x^\mu}{d\lambda^2} + \Gamma^{\mu}_{\alpha\nu}
\frac{dx^\alpha}{d\lambda}\frac{dx^\nu}{d\lambda} =
0\,,
\ee
where $\Gamma^{\mu}_{\alpha\nu}$ is the Christoffel symbol, and
$\lambda$ is a monotonically increasing (or decreasing) parameter
defined along the path of the photons.

Due to axial symmetry, the photon paths must be independent of the
azimuth angle $\phi$, which leaves three possible choices for free
index $\mu$. First, $\mu = t$ yields 
\be
\label{eq:du}
\frac{d^2 t}{d\lambda^2}+\frac{R'\dot{R}'}{1+\beta}
\left(\frac{dr}{d\lambda}\right)^2 + R\dot{R}
\left(\frac{d\theta}{d\lambda}\right)^2 = 0\,.
\ee 
Next, $\mu = r$ yields
\bean
\frac{d^2r}{d\lambda^2}
&+&\left(\frac{R''}{R'}-\frac{\beta'}{2+2\beta}\right)
\left(\frac{dr}{d\lambda}\right)^2 \nonumber \\ 
&+&\frac{2\dot{R}'}{R'}\frac{dr}{d\lambda}\frac{dt}{d\lambda}
-\frac{R(1+\beta)}{R'}\left(\frac{d\theta}{d\lambda}\right)^2 = 0\,,
\eean
and finally, $\mu = \theta$ yields
\be
\frac{d^2\theta}{d\lambda^2}
+2\frac{R'}{R}\frac{d\theta}{d\lambda}\frac{dr}{d\lambda}
+2\frac{\dot{R}}{R}\frac{d\theta}{d\lambda}\frac{dt}{d\lambda} = 0\,,
\ee
which can be written as conservation of angular momentum $J$
\be
\frac{d}{d\lambda}\left(R^2\frac{d\theta}{d\lambda}\right) \equiv
\frac{d}{d\lambda}J = 0 \,.
\ee

In addition, the 4-velocity identity, $u^\mu u_\mu = 0$ for photons,
leads to the constraint 
\be
\label{eq:4vel}
-\left(\frac{dt}{d\lambda}\right)^2+\frac{(R')^2}{1+\beta}
\left(\frac{dr}{d\lambda}\right)^2+\frac{J^2}{R^2} = 0 \,.
\ee

\begin{figure}
\begin{center}
\input{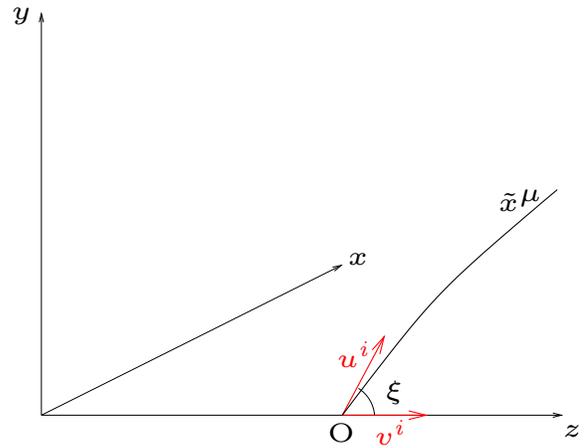}
\end{center}
\caption{A photon hitting the observer at an angle $\xi$} 
\label{fig:initcond}
\end{figure}

It is simplest to specify the initial conditions at the time $t_0$
when the  photon arrives at the observer's position, which is given by
$r=r_0$ and $\theta = 0$. The path of the photon is shown in
Fig.~\ref{fig:initcond}. It hits the observer at an angle $\xi$
relative to the $z$-axis. The spatial components of the unit vector
along this axis are
\be
v^i = \frac{\sqrt{1+\beta}}{R'}(1,0,0)\,,
\ee
where the three components are in the $r$, $\theta$ and $\phi$
direction, respectively.

The spatial direction $u^i$ is given by the tangent to the photon path
at 
$t_0$, i.e.
\be
u^i = \left|\frac{d\lambda}{dt}\right| 
\left(\frac{dr}{d\lambda},\frac{d\theta}{d\lambda},
  \frac{d\phi}{d\lambda}\right) = -\frac{1}{u}(p,J/R^2,0)\,,
\ee
where the first factor ensures normalization, $g_{ij}u^i u^j = 1$, and
we have introduced $u \equiv dt/d\lambda$ and $p \equiv dr/d\lambda$.
Note that we have chosen to let $\lambda$ decrease with time, since we
will start integrating the equations at $t=t_0$ and follow the photons
backwards in time until recombination.

The angle $\xi$ is given by the inner product of $v^i$ and $u^i$
\be
\label{eq:initalpha}
\cos \xi = g_{ij}u^i v^j = -\frac{R'}{\sqrt{1+\beta}}\frac{p}{u}\,.
\ee
Since the parametrization of the photon path in terms of the affine
parameter $\lambda$ is arbitrary, we can choose this such that
$\lambda=0$ when $t=t_0$ and $u_0=u(\lambda=0)=-1$. Using
Eqs.~(\ref{eq:4vel}) and (\ref{eq:initalpha}), these conditions
translate into the following initial conditions
\bean
p_0 &=& \frac{\sqrt{1+\beta}}{R'}\cos \xi \,,\\
\label{eq:Jinit}
J_0 &=& J = R\sin\xi\,.
\eean

Furthermore, we need to determine the redshift of the incoming photons
as a function of the direction. The photons follow a path given by
$r(\lambda)$, $t(\lambda)$ and $\theta(\lambda)$. The redshift can be represented by the change in time separation between adjacent photons. Due to the
expansion of the universe, this separation changes as the photons
propagate through space. The redshift is given by the relative change of the
separation, i.e. $z = (\tau_r-\tau_e)/\tau_e$, where the subscripts
$r$ and $e$ refer to the receiver and emitter positions, respectively. 

Consider two photons emitted by a source with a time separation of
$\tau$. Let the equation describing the time coordinate along the first
geodesic be $t_1(\lambda)=t(\lambda)$. The time along the second
geodesic is then given by $t_2(\lambda)=t(\lambda)+\tau(\lambda)$.
Both photons must satisfy the 4-velocity identity (\ref{eq:4vel}). For 
the first photon this reads as
\be
\label{eq:4v1}
\left(\frac{dt}{d\lambda}\right)^2 = \frac{R'(r,t)^2}{1+\beta}
\left(\frac{dr}{d\lambda}\right)^2 +
R(r,t)^2\left(\frac{d\theta}{d\lambda}\right)^2 ,
\ee
while for the second photon, the 4-velocity identity becomes
\be
\label{eq:4v2}
\begin{split}
\left(\frac{d(t+\tau)}{d\lambda}\right)^2 = &
\frac{R'(r,t+\tau)^2}{1+\beta} \left(\frac{dr}{d\lambda}\right)^2 \\
&+ R(r,t+\tau)^2\left(\frac{d\theta}{d\lambda}\right)^2.
\end{split}
\ee
Expanding Eq.~\eqref{eq:4v2} to first order in $\tau$ and using
Eq.~\eqref{eq:4v1}, we arrive at the expression
\be
\label{eq:diffT}
\frac{dt}{d\lambda}\frac{d\tau}{d\lambda} = \tau(\lambda) \left[
\frac{R'\dot{R}'}{1+\beta}\left( \frac{dr}{d\lambda} \right)^2 +
R\dot{R}\left( \frac{d\theta}{d\lambda} \right)^2 \right]\,.
\ee
The redshift measured by the observer as a function of the period at
the time of emission is defined as
\be
\label{eq:DefOfz}
1+z(\lambda_e)=\frac{\tau(\lambda_r)}{\tau(\lambda_e)}\,.
\ee
We differentiate this with respect to $\lambda_e$, which gives us the 
expression
\be
\frac{dz}{d\lambda_e}=-\frac{1}{\tau(\lambda_e)}
  \frac{d\tau(\lambda_e)}{d\lambda_e}
  \frac{\tau(\lambda_r)}{\tau(\lambda_e)} \,.
\ee
Next, using Eqs.~\eqref{eq:diffT} and \eqref{eq:DefOfz}, and
suppressing the subscript $e$, we arrive at the equation
\be
\frac{dz}{d\lambda} = -
(1+z)\frac{d\lambda}{dt}\left[\frac{R'\dot{R}'}{1+\beta}
  \left(\frac{dr}{d\lambda}\right)^2  +
  R\dot{R}\left(\frac{d\theta}{d\lambda}\right)^2\right]\,. 
\ee
This equation determines the change in redshift measured by the observer along an infinitesimal 
distance $d\lambda$. To find the redshift as a function of $\lambda$
for a photon hitting the observer today, we can sum up  the infinitesimal contributions
 along the past light cone, 
\be
\label{eq:dzdlambda}
\frac{d \ln (1+z)}{d\lambda} =
-u^{-1}\left[\frac{R'\dot{R}'}{1+\beta}p^2 +
  \frac{\dot{R}}{R^3}J^2\right] \,,
\ee
with the initial condition $z(\lambda=0) = z_0 = 0$. 

To summarize, we must solve the five first-order differential
equations for ($t, r, \theta, p$ and $z$), with the corresponding
initial conditions ($t_0, r_0, 0, p_0$ and $z_0$), under the
constraints~(\ref{eq:4vel}) and~(\ref{eq:Jinit}).


\section{Temperature anisotropies}
\label{sec:temp}
We wish to examine how being situated away from the center of the LTB
coordinate system affects the CMB temperature measured by the observer.
Since space-time is no longer spherically symmetric around such an
observer, we expect him to measure additional anisotropies in the
temperature to those measured by an observer at the center. In this
paper we concentrate on the additional anisotropies rising from the
observer's location, i.e. we disregard any intrinsic anisotropies in the
CMB temperature at the last-scattering surface. Thus, we assume the
temperature at the last-scattering surface to be isotropic. Any
anisotropies measured by observers today are therefore due to the
propagation of photons through an anisotropic space-time.

The temperature of the background radiation in a given direction is determined by measuring the intensity of incident photons from this direction. Assuming the radiation to be black-body radiation, the intensity will be given by a Planck spectrum with a corresponding characteristic temperature. It can be shown that the radiation field preserves its black-body nature as it propagates freely through space under the influence of a gravitational field \cite{Ellis:1971}. At any later time, the spectrum will still remain a Planck spectrum, but with a different temperature. 

In our specific case, the CMB temperature seen today by an off-center observer is given by 
\be
\label{eq:defTemp}
T(\xi)=\frac{T_*}{1+z(\xi)}\,,
\ee
where $T_*$ is the temperature at the last-scattering surface, and $\xi$ is the angle defined in Fig.~\ref{fig:initcond}. The average temperature $\widehat{T}$
measured by the observer is then
\be
\label{eq:defTnow}
\widehat{T}\equiv\frac{1}{4\pi}\int\!d\Omega\, T(\xi) = 
\frac{T_*}{2}\int_0^\pi d\xi\frac{\sin{\xi}}{1+z(\xi)}\,.
\ee
According to measurements made by the COBE satellite
\cite{Mather:1998gm}, this temperature is $\widehat{T}=2.725$. We can now
use Eqs.~\eqref{eq:defTemp} and \eqref{eq:defTnow} to define an
average redshift to the last-scattering surface: 
\be
\label{eq:defzLSS}
1+z_{*}\equiv\frac{T_*}{\widehat{T}}=2\left[\int_0^\pi\!d\xi
\frac{\sin{\xi}}{1+z(\xi)}\right]^{-1}\,.
\ee
The relative temperature variation measured by the observer today will
then be
\be
\label{eq:Theta}
\Theta(\xi)\equiv\frac{\Delta T}{\widehat{T}}=\frac{T(\xi)-
  \widehat{T}}{\widehat{T}}= \frac{z_{*}-z(\xi)}{1+z(\xi)}\,.
\ee
It is often more interesting to consider contributions at different
angular scales rather than the total anisotropy itself. Such an
analysis can be performed by decomposing the temperature field in
spherical harmonics $Y_{lm}$: 
\be
\label{eq:ThetaSH}
\Theta(\xi) = \sum_{l,m}a_{lm}Y_{lm}\,,
\ee
where the amplitudes in the expansion are recovered as
\be
\label{eq:alm}
a_{lm}=
\int_0^{2\pi}\!\int_0^\pi\!\Theta\,Y_{lm}^* \sin{\xi}\, d\xi\,
d\phi \,.
\ee
These measure the level of anisotropy at different angular
scales, with larger $l$ values corresponding to smaller scales. Since
the relative temperature field does not depend on the azimuth angle
$\phi$, all the $a_{lm}$ will vanish, except those with $m=0$. 

The observed dipole in the CMB is of the order $|a_{10}|\sim10^{-3}$.
This will put a natural constraint on how far away from the origin the 
observer can be located, since a farther off-center position usually 
means a larger dipole.

\section{Numerical results}
\label{sec:numeric}

Following \cite{alnes:2005} we will consider LTB models where the inhomogeneity is an underdense bubble with a flat and homogeneous space-time outside. We will consider two specific models of this type corresponding to two different choices of the functions $\alpha(r)$ and $\beta(r)$ in Eqs.~\eqref{eq:defR} and \eqref{eq:deft}. We will refer to these as \emph{model I} and \emph{model II}. We parametrize these function in the same way as in \cite{alnes:2005}, i.e. we write 
\bean
\alpha(r) &=& H_{0}^2r^3\left[ \alpha_0-\Delta \alpha
  \left(\frac{1}{2}-\frac{1}{2}\tanh \frac{r-r_0}{2\Delta r}\right)
  \right]\\  
\beta(r) &=& H_{0}^2r^2\left[ \beta_0 - \Delta \beta 
  \left(\frac{1}{2}-\frac{1}{2}\tanh \frac{r-r_0}{2\Delta r}\right)
  \right]
\eean
which corresponds to a smooth interpolation between two homogeneous
regions (i.e. a spherical bubble in an otherwise homogeneous
universe). The parameter $H_{0} = 100\, h_{\text{out}} \unit{km\, s^{-1}
  Mpc^{-1}}$ is the Hubble constant of the outer homogeneous region
today, while $\alpha_0$ and $\beta_0 = 1-\alpha_0$ are the relative
densities of matter and curvature in this region. Further, $\Delta
\alpha = -\Delta \beta$ determines the difference in matter density
between the regions, while $r_0$ and $\Delta r$ specify the position
and width of the transition. 

\begin{figure}
\begin{center}
\includegraphics[width=8cm]{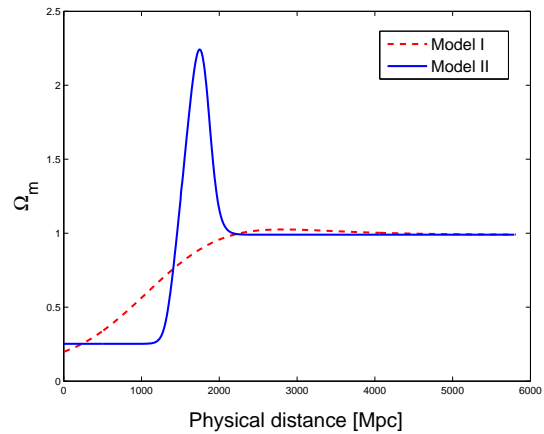}
\end{center}
\caption{The matter density today as a function of physical distance
  for the two models considered.}
\label{fig:O_m}
\end{figure}

In our previous work \cite{alnes:2005}, we assumed that the observer was 
positioned at the center of the bubble, and found a model that gave a
good agreement with the Hubble diagram of observed SNIa and the
position of the first CMB peak. Model I is identical to this model,
while model II is slightly different. The matter distribution today for
these models is plotted in Fig.~\ref{fig:O_m}, where we have used the generalized matter density defined in \cite{alnes:2005}. Various other
properties of these are listed in Tab.~\ref{tab:models}. One notable
difference between these two models is that the transition from the
underdensity to the homogeneous region is much sharper in the second
model. Note that the physical values given in the table are found
assuming that the observer is placed at the center. The shift
parameter $\mathcal{S}$ is defined in \cite{alnes:2005}, and is simply
the shift of the first peak in the CMB power spectrum relative to the
concordance $\Lambda$CDM model. As we can see, both models yield a
very good fit to both SNIa and the first CMB peak. 

\squeezetable
\begin{table}[t]
\begin{tabular}{lccc}
\hline 
Description & Symbol & Model I & Model II\\
\hline 
Density contrast parameter & $\Delta \alpha$         & 0.90  & 0.78 \\
Transition point [Gpc] & $d_0$                       & 1.34  & 1.68 \\
Transition width & $\Delta d/d_0$                    & 0.40  & 0.03 \\
Fit to supernovae & $\chi^2_{SN}$                    & 176.2 & 177.8\\
Position of first CMB peak & $\mathcal{S}$           & 1.006 & 1.002\\
Age of the universe [Gyr] & $t_0$                    & 12.8  & 12.7\\
Relative density at the center & $\Omega_{m,in}$ & 0.20  & 0.25\\
Relative density outside underdensity & $\Omega_{m,out}$ & 1.00 & 1.00\\
Hubble parameter at the center & $h_{in}$ & 0.65  & 0.63\\
Hubble parameter outside underdensity & $h_{out}$ & 0.51 & 0.51\\
\hline
\end{tabular}
\caption{The parameters and features of two 
  inhomogeneous models, with the observer placed at the center. Note
  that $d_0$ and $\Delta d$ are physical distances corresponding to
  $r_0$ and $\Delta r$}
\label{tab:models}
\end{table}

\begin{figure}
\begin{center}
\includegraphics[width=7.0cm]{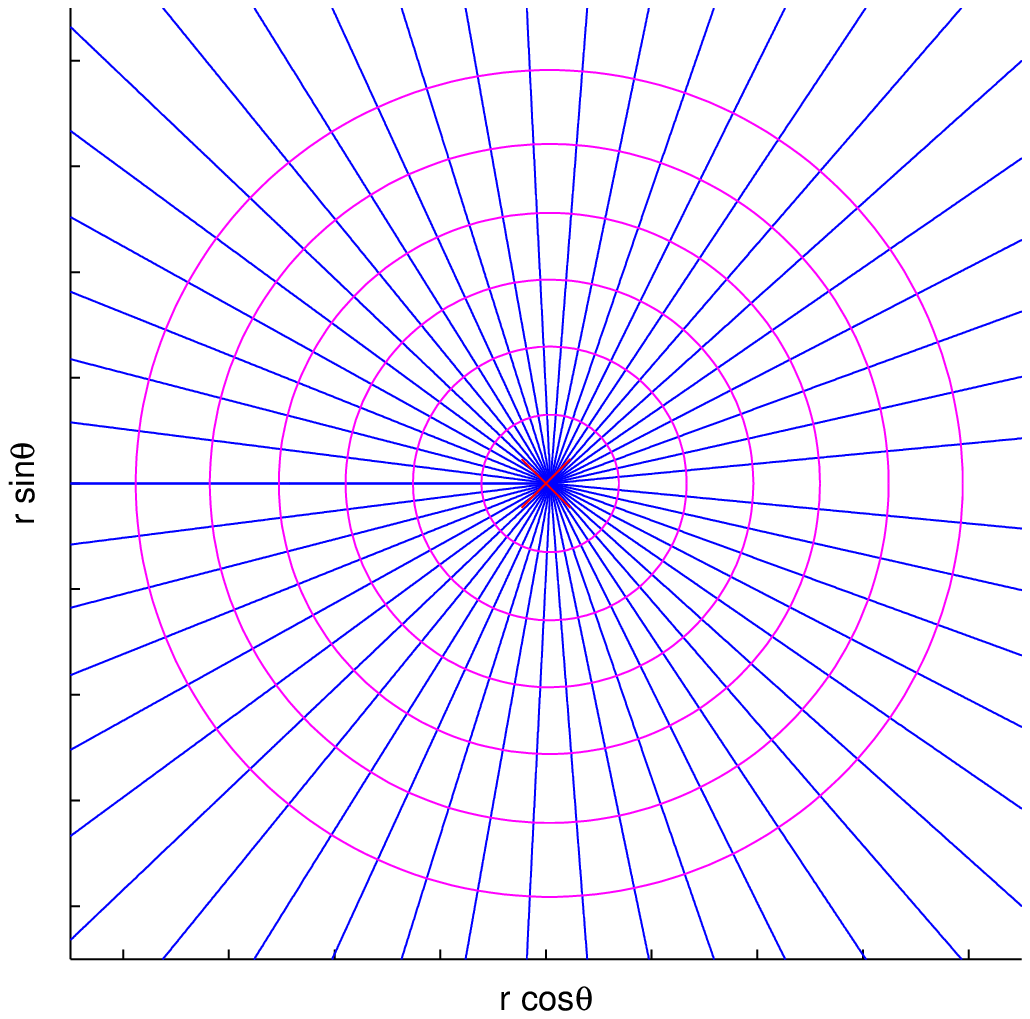}\\
\includegraphics[width=7.0cm]{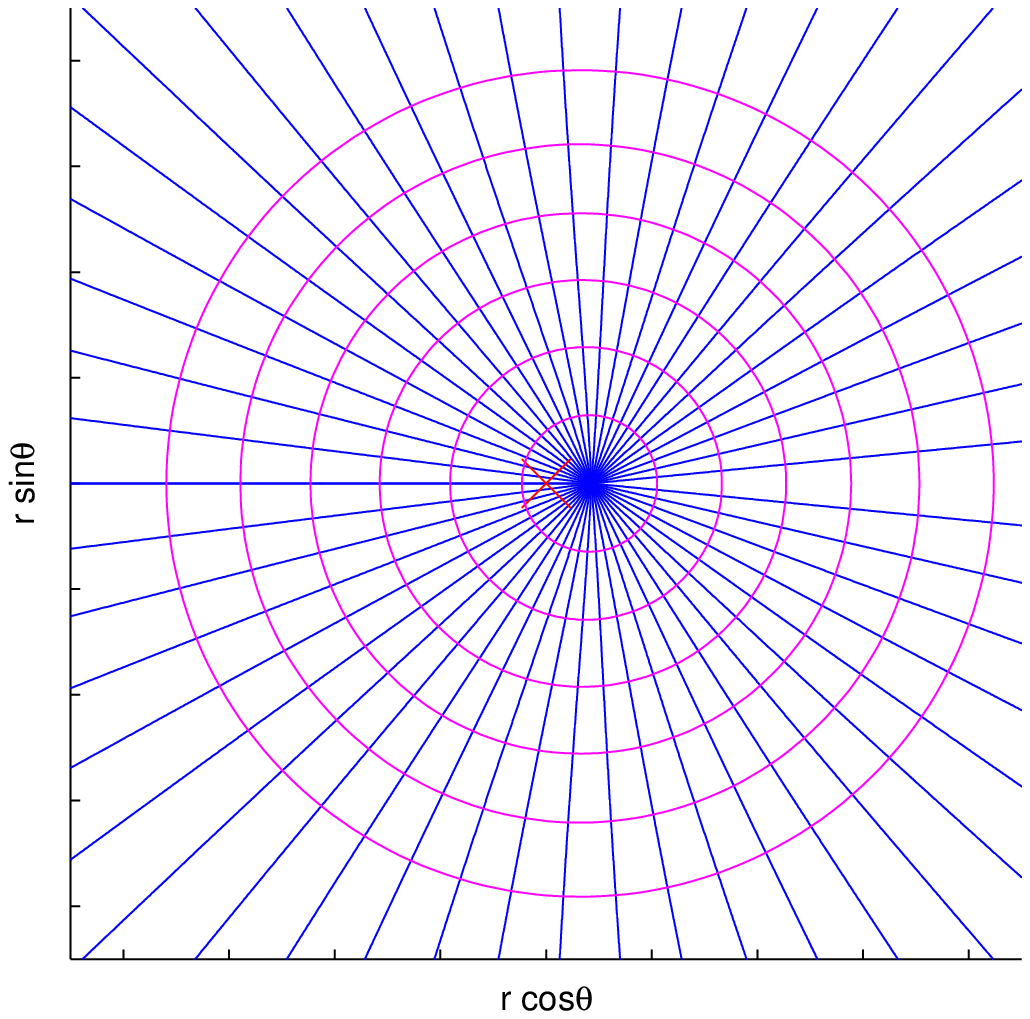}
\end{center}
\caption{Model I: Photon paths converging on the observer located
 $\unit[20]{Mpc}$ (top) and $\unit[200]{Mpc}$ (bottom) from the
 center. The red cross marks the center of the underdensity, while the
 red circles show positions that are evenly spaced in cosmic time with
 a separation of $\unit[1]{Gyr}$
} 
\label{fig:geodets1}
\end{figure}

\begin{figure}
\begin{center}
\includegraphics[width=7.0cm]{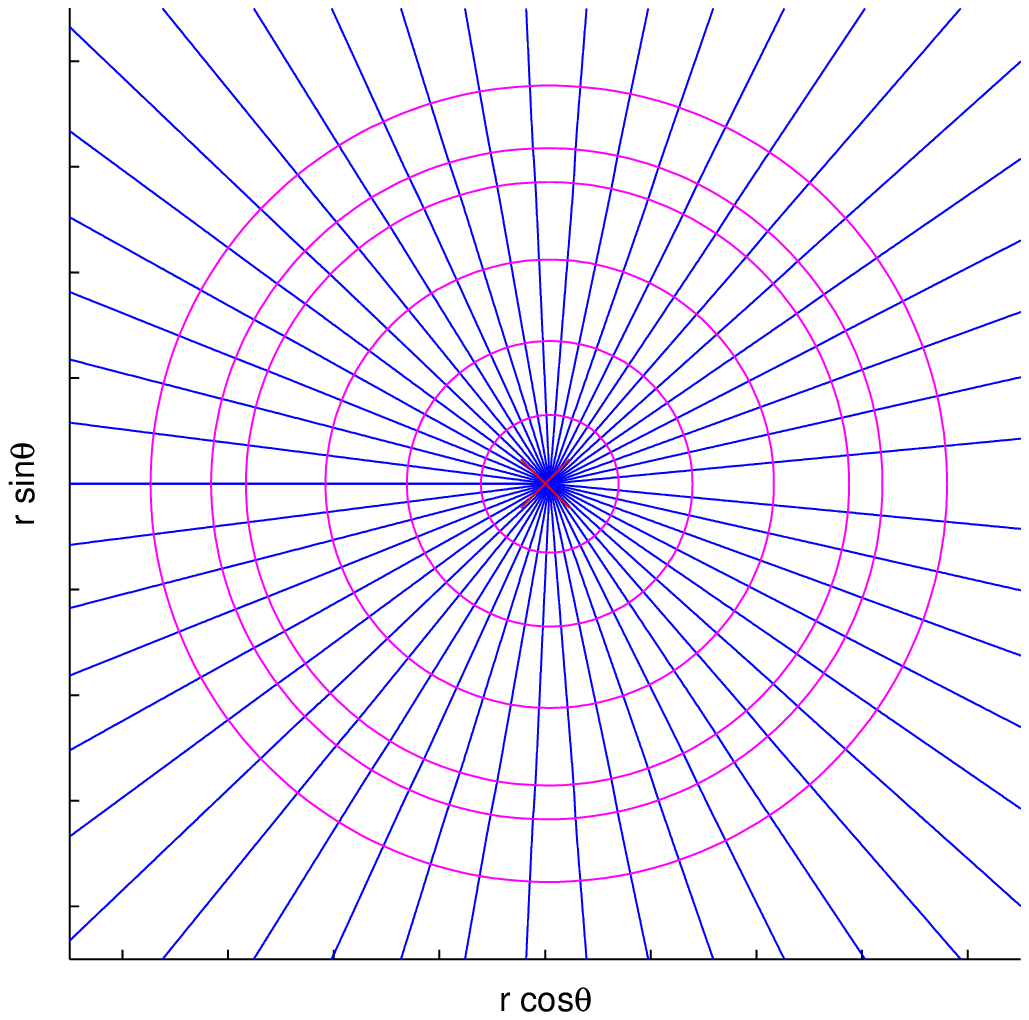}\\
\includegraphics[width=7.0cm]{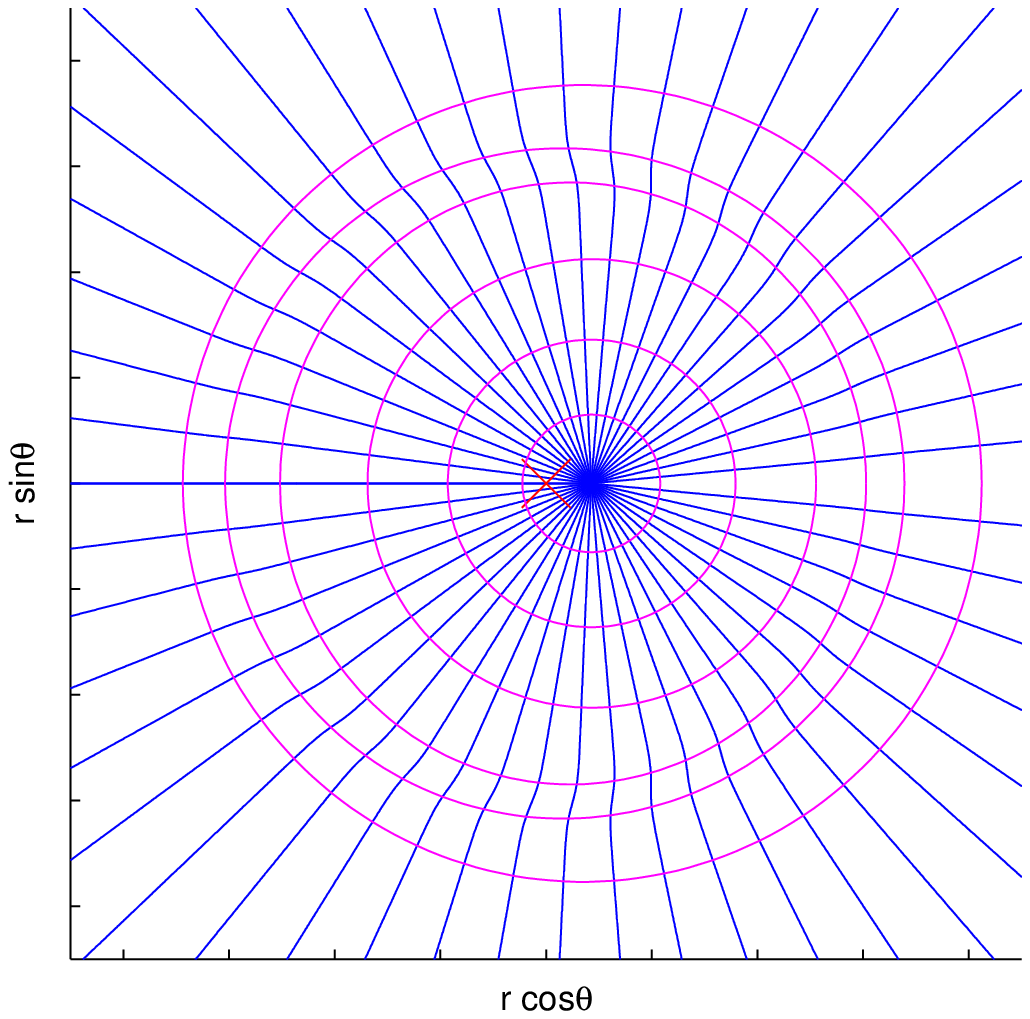}
\end{center}
\caption{Model II: Photon paths converging on the observer located $\unit[20]{Mpc}$ (top) and $\unit[200]{Mpc}$ (bottom) from the center. The red cross marks the center of the underdensity, while the
 red circles show positions that are evenly spaced in cosmic time with
 a separation of $\unit[1]{Gyr}$}
\label{fig:geodets2}
\end{figure}

Figs.~\ref{fig:geodets1} and \ref{fig:geodets2} show the solutions of
the geodesic equations for two off-center observers in the two models,
with the observers located $\unit[20]{Mpc}$ and $\unit[200]{Mpc}$ from
the center. The blue lines show the photon paths in the metric
($r,\theta$) space, while the red circles are the positions of
incoming photons at evenly spaced points in (cosmic) time.
The distortion of light paths is clearly visible on the bottom plot of
Fig.~\ref{fig:geodets2}, in which the observer is placed relatively
far from the center of the inhomogeneity and the density gradient is
large at the transition.

In our calculations, recombination is assumed to occur at $t=0$ and
today ($t_0$) is defined to be the time when the redshift of photons
emitted at $t=0$ reaches $z_* \simeq 1100$. (The exact value of $z_*$
depends on the matter density outside the bubble, and a fitting
formula is given in \cite{Doran:2001yw}).  

\begin{figure*}
\begin{center}
\includegraphics[width=6.0cm]{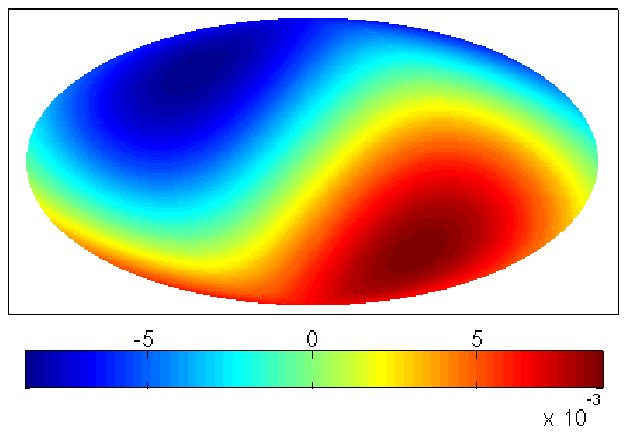}
\includegraphics[width=6.0cm]{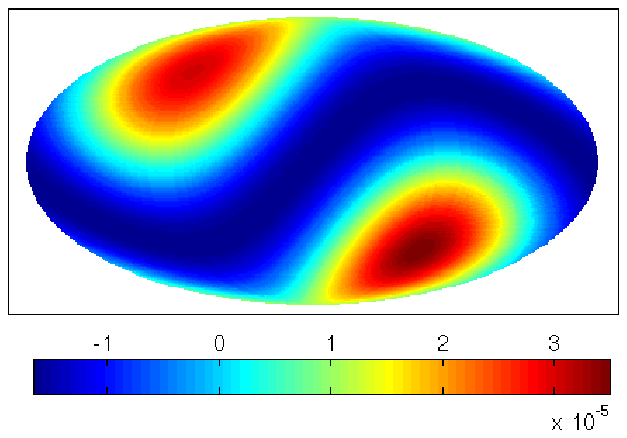}
\includegraphics[width=6.0cm]{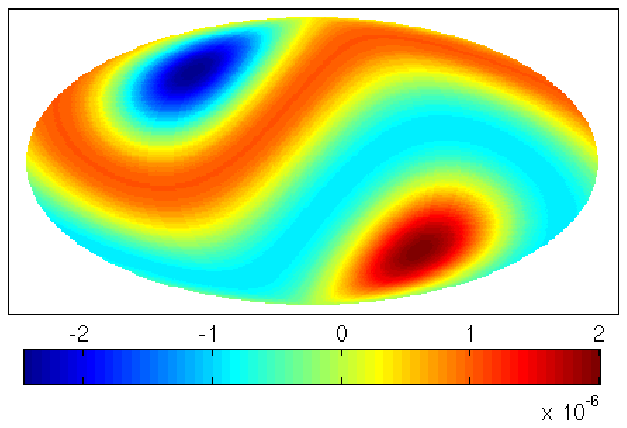}
\includegraphics[width=6.0cm]{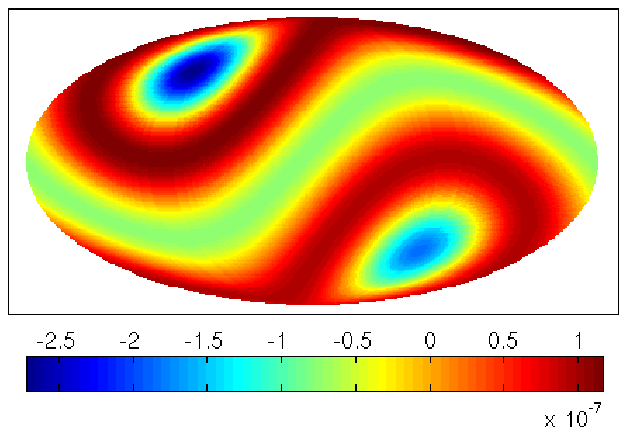}
\end{center}
\caption{Temperature anisotropy in galactic coordinates seen by an observer $\unit[200]{Mpc}$
  from the center of Model I, oriented so that the induced dipole
  coincides with the direction of the dipole seen by the COBE
  satellite \cite{Mather:1998gm}. The top plot to the left shows the temperature
  map, which is completely dominated by the dipole. Then follow plots
  with the dipole, quadrupole and octopole removed, successively}
\label{fig:tempplot}
\end{figure*}

As discussed in the previous section, an off-center observer will
measure a temperature anisotropy due to the non-symmetric paths
traversed by CMB photons in different direction in the sky. Using
Eqs.~\eqref{eq:ThetaSH} and \eqref{eq:alm}, we can now calculate the 
temperature multipoles seen by such an observer. As an example, a plot
of the multipoles can be seen in Fig.~\ref{fig:tempplot} for an
observer who is located $\unit[200]{Mpc}$ from the center in model I.

\begin{figure*}
\begin{center}
\includegraphics[width=5.5cm]{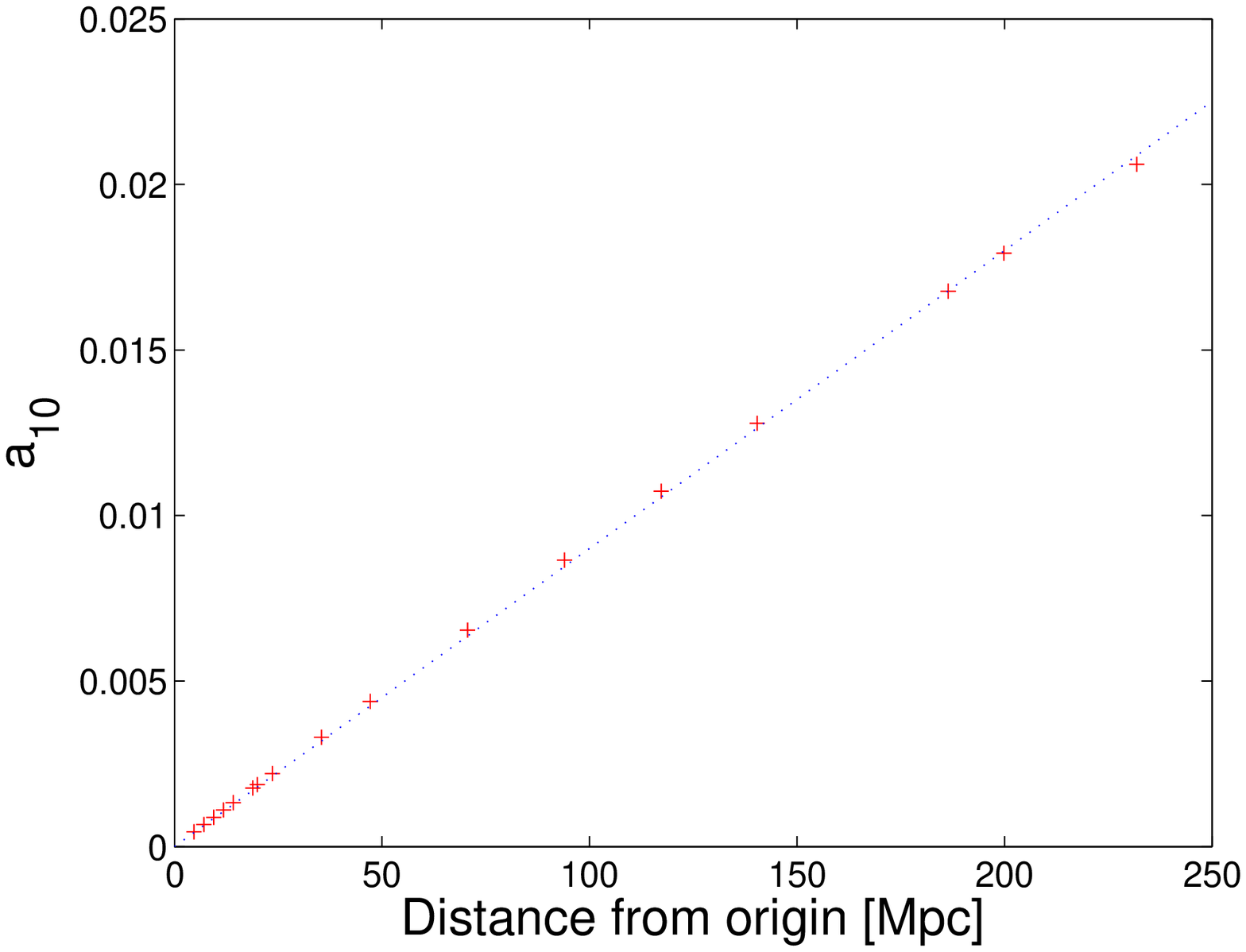}
\includegraphics[width=5.5cm]{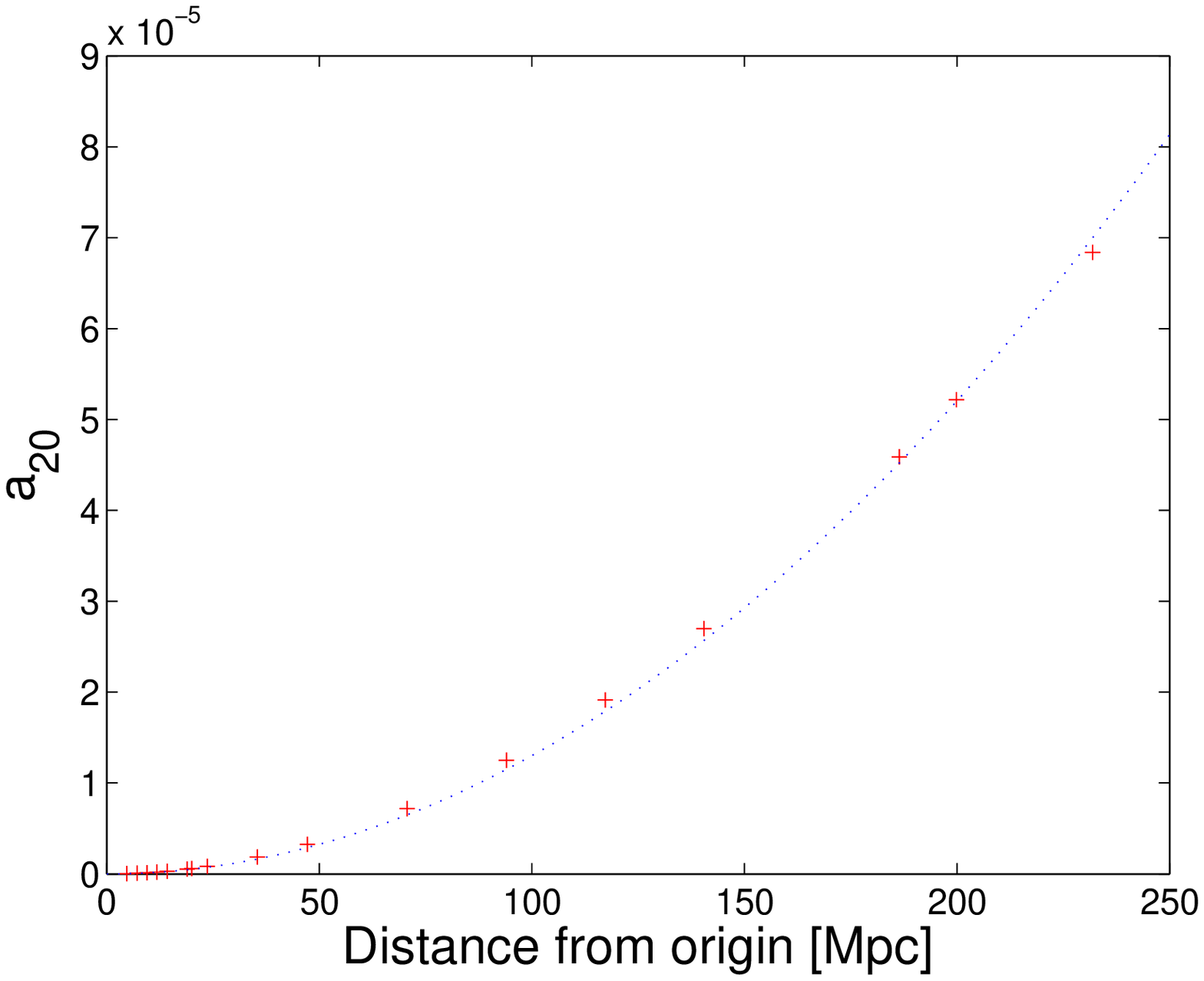}
\includegraphics[width=5.5cm]{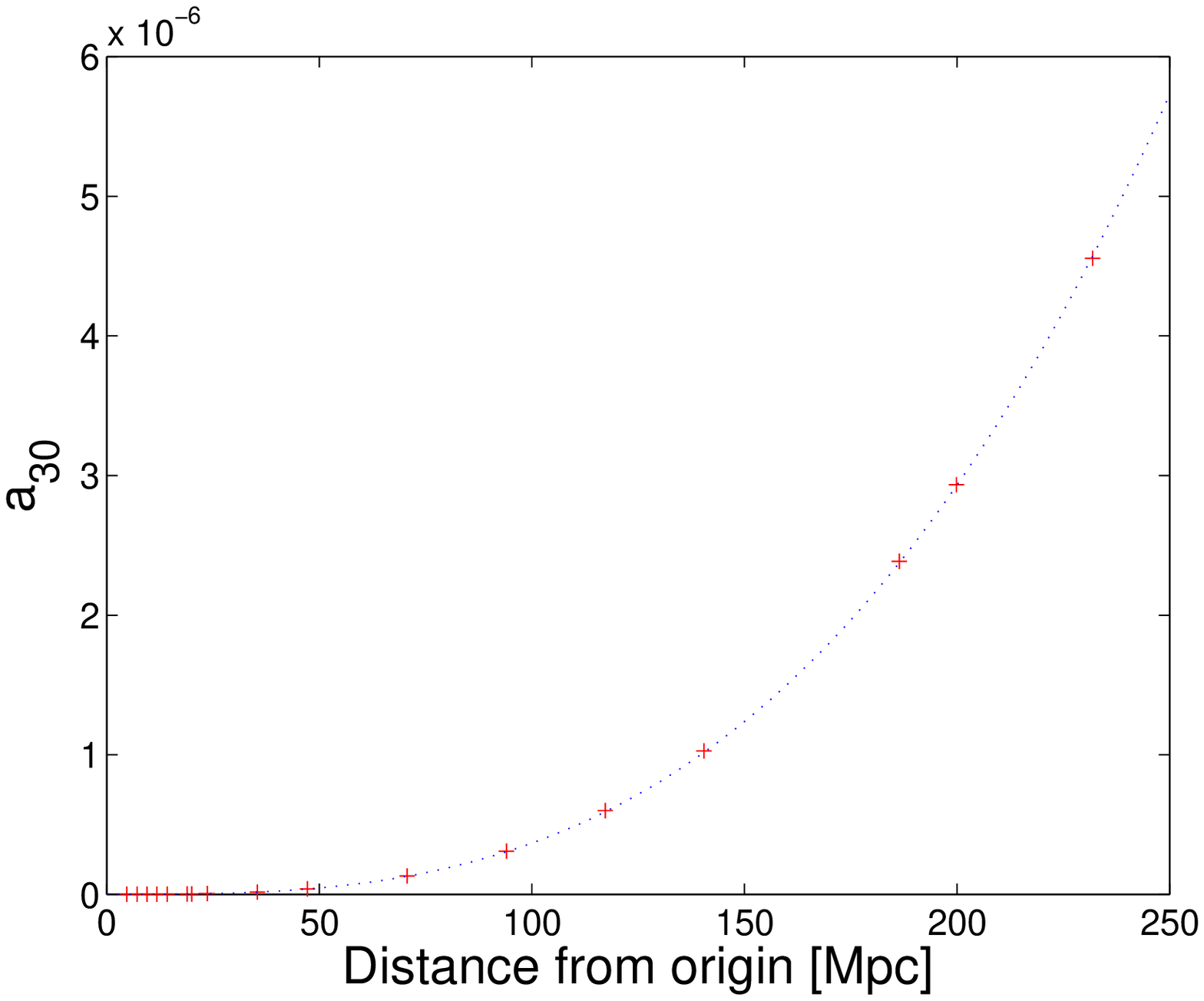}
\end{center}
\caption{The $a_{l0}$ as a function of the observer's position, in
  Model I. The dotted lines are linear, quadratic and cubic fits, respectively.} 
\label{fig:alms_1}
\end{figure*}

In Fig.~\ref{fig:alms_1}, the coefficients $a_{l0}$ for the dipole
($l=1$), quadrupole ($l=2$) and octopole ($l=3$) are plotted as functions of the observer's position in model I.
The most striking feature of these plots is that the quadru- and octopoles are very small compared to the dipole. If we assume that the induced dipole must be smaller than $10^{-3}$, the induced quadrupole is less than $10^{-7}$ while the induced octopole is smaller than $10^{-9}$. 

\begin{figure*}
\begin{center}
\includegraphics[width=5.5cm]{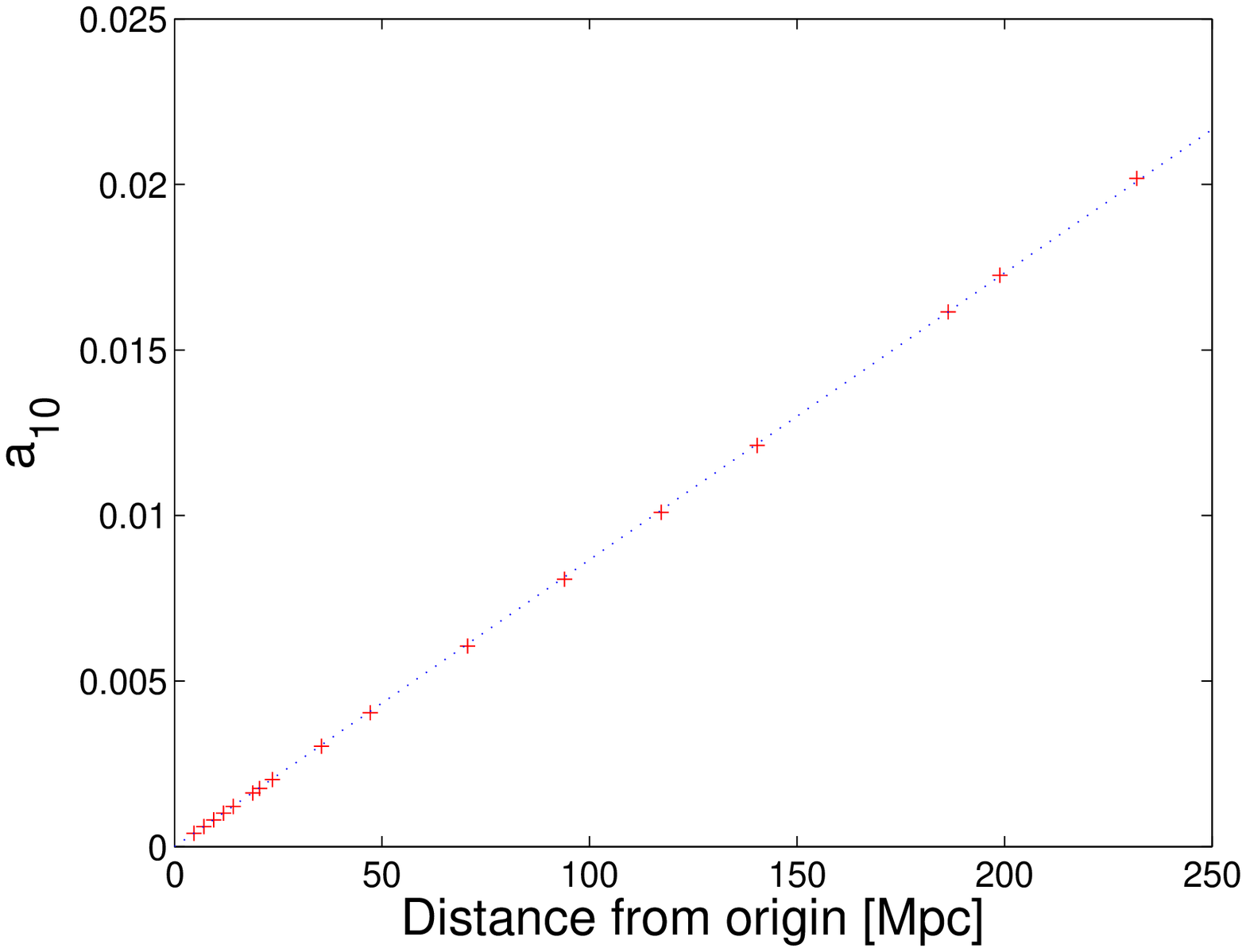}
\includegraphics[width=5.5cm]{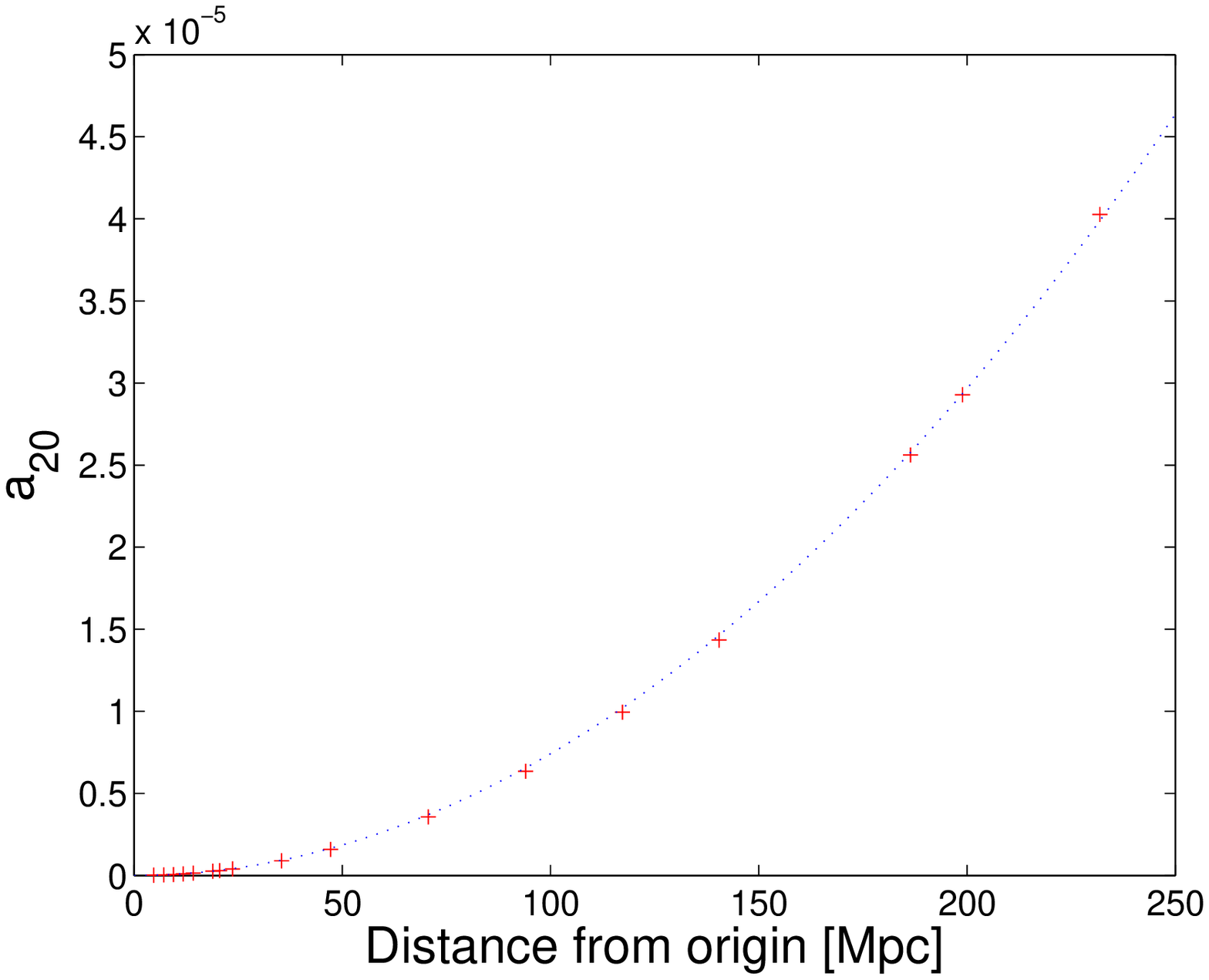}
\includegraphics[width=5.5cm]{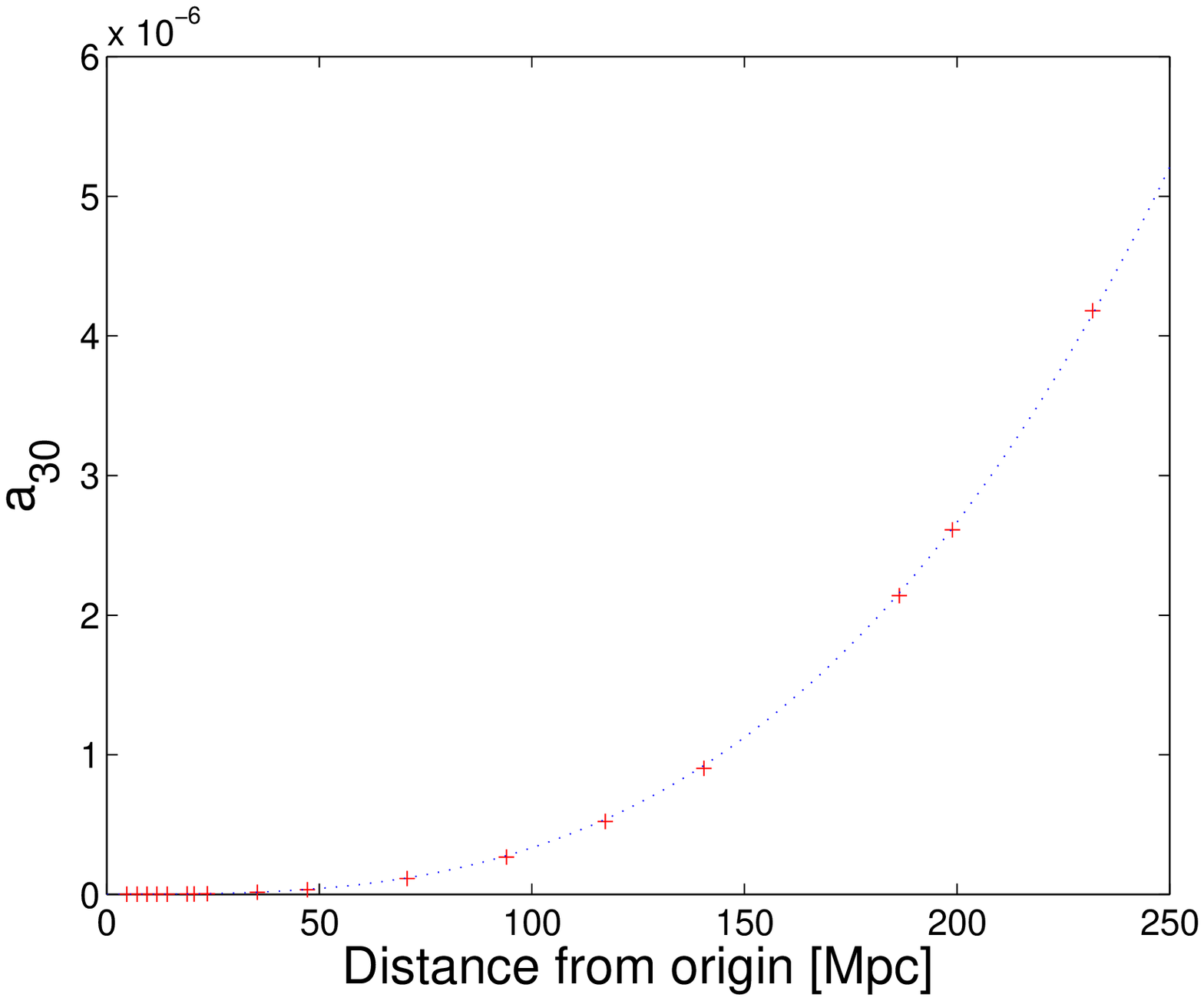}
\end{center}
\caption{The $a_{l0}$ as a function of the observer's position, in
  Model II. The dotted lines are linear, quadratic and cubic fits, respectively.} 
\label{fig:alms_2}
\end{figure*}

In Fig.~\ref{fig:alms_2}, the $a_{l0}$'s of model II are plotted as functions of the observer's position.  
It is evident that the behavior is almost indistinguishable from that 
of Model I, except for the largest distances where the transition from 
underdensity to flat space is starting to show in the first model.

\section{Discussion}
\label{sec:discussion}
The main purpose of this paper has been to determine the maximum displacement of the  observer from the origin of the underdensity, for which the induced CMB dipole remains in  agreement with the results observed by COBE \cite{Bennett:1996ce}. Of course, one could in principle introduce an additional peculiar velocity towards the center of the underdensity to compensate for a too large induced dipole, but such a coincidence would be very  difficult to justify. Therefore, we must require the induced $a_{10}$ to be of order $10^{-3}$ or less, which from the plots in Figs.~\ref{fig:alms_1} and \ref{fig:alms_2} can be translated to
\be
\label{eq:r_requirement}
d_{obs} \lesssim \unit[15]{Mpc}\,.
\ee
where $d$ is the physical distance. When compared to the size of the underdensity, which according to Fig.~\ref{fig:O_m} is around \unit[1\,500]{Mpc}, this means that if we are placed at a random position inside the
bubble, there is roughly a chance of $1$ to $10^6$ that we end up inside the region allowed by Eq.~\eqref{eq:r_requirement}. This is a rather strong violation of the Copernican principle, which states that we are \emph{not} situated at a special place in the universe. On the other hand, a $10^{-6}$ probability is still much better than the  infinitely improbable case of the observer being \emph{exactly} at the center of the  underdensity. Note that the size of the underdensity is dictated by the fit to the CMB and SNIa data. We have not been able to find smaller bubbles that fit these data as
well as the models considered here. 

From Figs.~\ref{fig:alms_1} and \ref{fig:alms_2} we see that the induced multipoles become larger the farther away from the origin the observer is located, as we would expect. Thus, the largest possible quadru- and octopoles with a dipole compatible with COBE measurements are those for an observer about $\unit[15]{Mpc}$ from the origin. However, at this relatively small distance, the values for these are of the order $10^{-7}$ for the quadrupole, and $10^{-9}$ for the octopole. It is therefore clear that the induced quadru- and octopole cannot explain the observed alignment of the low-l multipoles in the CMB, since their contributions are negligible compared to the observed anisotropies (which are of order $10^{-5}$). Furthermore, any off-center placement must necessarily result in axial
symmetric contributions to the CMB spectrum. Even if such contributions were of the correct order, Raki\'{c} et al. \cite{Rakic:2006tp} show that they are very unlikely to explain the alignment.

The smallness of the induced multipoles can be understood from a simplified Newtonian picture. Compared to the homogeneous case with a spatially constant Hubble parameter $h_{out}$, the observer at $d_{obs}$ has a ``peculiar velocity'' of roughly
\be
\beta = \frac{v_p}{c} = \frac{h_{in}-h_{out}}{\unit[3000]{Mpc}}d_{obs}
\ee
with respect to the origin. In such a picture, the temperature anisotropies measured by the observer are attributed to a Doppler shift of the CMB photons due to his motion. The change in frequency can then be written as \cite{jackson}
\be
\label{eq:nushift}
\frac{\nu_o}{\nu_e}=\frac{\sqrt{1-\beta^2}}{1-\beta\cos{\xi}}\,,
\ee
where $\nu_o$ is the frequency measured by the observer, and $\nu_e$ is the frequency relative to a stationary background. The temperature shift associated with this frequency shift of the CMB photons is
\be
\label{eq:TshiftNewton}
\frac{T_o}{T_e}=\frac{\sqrt{1-\beta^2}}{1-\beta\cos{\xi}}\,.
\ee
The average background temperature measured by the observer will then be
\be
\label{eq:AvT}
\widehat{T}_o=\frac{1}{4\pi}\!\!\int d\,\Omega T_o(\xi) = \frac{T_e}{2} \frac{\sqrt{1-\beta^2}}{\beta}\ln{\left(\frac{1+\beta}{1-\beta}\right)}\,.
\ee
Thus, the temperature anisotropy can be written as
\be
\label{eq:dT}
\Theta(\xi) = \frac{\Delta T(\xi)}{\widehat{T}_o} = \frac{2}{\ln{\left( \frac{1+\beta}{1-\beta} \right)}} \frac{\beta}{1-\beta\cos{\xi}}-1\,.
\ee
Using Eq.~\eqref{eq:alm}, the multipoles can now be calculated in terms of the velocity of the observer. We find that the leading contribution to $a_{l0}$ is $\beta^l$,
\be
\label{eq:newtonian}
a_{l0} \sim \beta^l \sim d_{obs}^l\,.
\ee
Thus, using this simplified Newtonian picture, we expect the dipole to scale linearly, the quadrupole quadratically, and the octopole cubically with the observers position. In Figs.~\ref{fig:alms_1} and \ref{fig:alms_2} we have plotted the real dependence of these multipoles on the observers position along with a best-fit $d_{obs}^l$ dependence. As we see, the Newtonian picture gives a very good description. The expressions for the three lowest multipoles to the lowest order in $\beta$ is
\bean
a_{10}=&\sqrt{\frac{4\pi}{3}}\frac{h_{in}-h_{out}}{\unit[3000]{Mpc}}d_{obs} \label{a10_newt} \,,\\
a_{20}=&\sqrt{\frac{16\pi}{45}}\left(\frac{h_{in}-h_{out}}{\unit[3000]{Mpc}}\right)^2d_{obs}^2 \label{a20_newt}\,, \\
a_{30}=&\sqrt{\frac{16\pi}{175}}\left(\frac{h_{in}-h_{out}}{\unit[3000]{Mpc}}\right)^3d_{obs}^3 \label{a30_newt}\,.
\eean
Numerically, this approximation yields $a_{10} = 1.4\times10^{-3}$, $a_{20} = 5.2\times10^{-7}$ and $a_{30} = 1.2\times10^{-9}$ for an observer at $d_{obs} = \unit[15]{Mpc}$ in Model I, whereas the exact values are $a_{10} = 1.4\times10^{-3}$, $a_{20} = 2.3\times10^{-7}$ and $a_{30} = 1.3\times10^{-9}$ respectively.  

Eqs.~\eqref{a10_newt}-\eqref{a30_newt} imply that it is impossible to obtain sufficiently large values for the quadru- and octopole as long as the dipole is within the limits set by the COBE data. Note, however, that Tomita \cite{Tomita:1999rw} has previously found relatively large values for the quadrupole in more simplified bubble models (where two homogeneous regions are separated by a massive comoving shell). It is unclear to us why the Newtonian approach fails for his models. We have attempted to reproduce his results by introducing very narrow transition regions in our continuous model, but the results we get for the multipoles are of the same order as those quoted above. We therefore expect Eq.~\eqref{eq:newtonian} to be roughly correct for all models of our type.

In our analysis so far we have only considered contributions to the multipoles from the off-center placement. There will of course be additional contributions from various sources such as the intrinsic primordial temperature anisotropies, the integrated Sachs-Wolfe (ISW) effect \cite{Sachs:1967} and a non-vanishing peculiar velocity of the observer. We have seen that when the dipole is constrained by data, the quadru- and octopoles due to the off-center placement are considerably weaker than those observed in the CMB. A possible way to obtain stronger quadru- and dipoles is to place the observer farther away from the center, while allowing one or more of the effects mentioned above to cancel out the excessive contribution to the dipole. 

However, concerning the first two effects, it is clear that neither of these can achieve such cancellation. Although there is no way of measuring directly the intrinsic dipole, it is reasonable to assume that it is of the same order as the neighboring multipoles, which are of order $10^{-5}$ Similarly, we expect the contribution to the dipole from the ISW effect to be of the same order as for the quadru- and octopole. Therefore, it is very unlikely that these effects are responsible for a chance cancellation of an excessive contribution to the dipole from the off-center placement.

A non-vanishing peculiar velocity can reduce the dipole to any desired value as long as the velocity is chosen large enough. However, multipoles due to such motion will have a hierarchical scaling similar to that which we showed in the Newtonian case. Thus, even if we manage to obtain values for the dipole and quadrupole of the correct order, the octopole would still be too weak. From this we can conclude that even when combined with other effects, the off-center placement cannot provide sufficient power to both the quadru- and octopole. 

In summary, LTB models like the ones listed in Table~\ref{tab:models}
are not ruled out on the basis of these results, but they do require a
violation of the Copernican principle, since the observer would have
to be located at a very special place. The volume within which the observer can be located is severely constrained by the size of the dipole induced by an off-center placement of the observer. As a consequence of this, the quadru- and octopole turn out to have insufficient power to explain the observed alignment.
However, the LTB models remain an exotic alternative to dark energy as
an explanation of the apparent accelerated expansion of the universe.

\begin{acknowledgments}
MA acknowledges support from the Norwegian Research Council through
the project "Shedding Light on Dark Energy" (grant 159637/V30). The
authors wish to thank Mike Hudson for a helpful correspondence.
\end{acknowledgments}

\bibliography{anis_061206.bbl}
   
\end{document}